\documentclass[letter,twocolumn]{cflowproc}

\usepackage{natbib}

\begin{document}

%%%%%%%%%%%%%%%%%%%%%%%%%%%%%%%%%%%%%%%%%%%%%%%%%%%%%%%%%%%%%%%%%%%%%%%%%%%%%%%%
%%
%% This section contains the author information and the title of your article
%%
%%%%%%%%%%%%%%%%%%%%%%%%%%%%%%%%%%%%%%%%%%%%%%%%%%%%%%%%%%%%%%%%%%%%%%%%%%%%%%%%

\shortauthors{J. O. Burns et al. }     % Use "et al." if more than 3 authors
\shorttitle{Non-Flowing Cool Cores} % Keep this under 40 characters

\title{On the Formation of Cool, Non-Flowing Cores in Galaxy Clusters via
Hierarchical Mergers}   % Replace this with your title

\author{Jack O. Burns,\affilmark{1}   % Put all authors in a single \author{} command
Patrick M. Motl,\affilmark{1}                % with \affilmark{}'s numbered as necessary 
Michael L. Norman\affilmark{2}
and Greg L. Bryan\affilmark{3}}

\affil{1}{University of Colorado, Center for Astrophysics and Space Astronomy, Boulder, CO 80309}   % affiliation numbers should match the affilmarks
\and                                % put an "\and" between each \affil{}
\affil{2}{University of California San Diego, Center for Astrophysics and Space Sciences, 9500 Gilman Drive, La Jolla, CA 92093}
\and
\affil{3}{University of Oxford, Astrophysics, Keble Road, Oxford OX1 3RH}

%%%%%%%%%%%%%%%%%%%%%%%%%%%%%%%%%%%%%%%%%%%%%%%%%%%%%%%%%%%%%%%%%%%%%%%%%%%%%%%%
%%
%% This section contains the abstract of your article
%%
%%%%%%%%%%%%%%%%%%%%%%%%%%%%%%%%%%%%%%%%%%%%%%%%%%%%%%%%%%%%%%%%%%%%%%%%%%%%%%%%

\begin{abstract}
We  present a new model for the creation of cool cores in rich galaxy clusters within a {$\Lambda$}CDM cosmological
framework using the results from high spatial dynamic range, adaptive mesh hydro/N-body simulations.  It is 
proposed that
cores of cool gas first form in subclusters and these subclusters merge to create rich clusters with cool, central X-Ray
excesses.  The rich cool clusters do not possess ``cooling flows'' due to the presence of bulk velocities in the
intracluster medium in excess of 1000 km/sec produced by on-going accretion of gas from supercluster filaments.  This new
model has several attractive features including the presence of substantial core substructure within the cool cores, and
it predicts the appearance of cool bullets, cool fronts, and cool filaments all of which have been recently observed with
X-Ray satellites.  This hierarchical formation model is also consistent with the observation that cool cores in Abell
clusters occur preferentially in dense supercluster environments.  On the other hand, our simulations overproduce cool
cores in virtually all of our numerical clusters, the central densities are high, and physical core temperatures are
often below 1 keV (in contrast to recent observations).  We will discuss additional preliminary simulations to ``soften''
the cool cores involving star formation and supernova feedback.
\end{abstract}

%%%%%%%%%%%%%%%%%%%%%%%%%%%%%%%%%%%%%%%%%%%%%%%%%%%%%%%%%%%%%%%%%%%%%%%%%%%%%%%%
%%
%% The body of the article starts here
%%
%%%%%%%%%%%%%%%%%%%%%%%%%%%%%%%%%%%%%%%%%%%%%%%%%%%%%%%%%%%%%%%%%%%%%%%%%%%%%%%%

%% Please start all labels with your last name and a colon to distinguish them
%% from labels that other authors may use in their proceedings contributions.
\section{Adaptive Mesh Refinement (AMR) Simulations of Cluster Formation and Evolution}
\label{Burns:intro}

In this paper, we  present numerical results from the simulation of the formation and evolution of
clusters of galaxies in the larger, cosmological context.  Our simulations are performed
with a sophisticated code that couples an N-body algorithm for evolving the collisionless
dark matter particles with an Eulerian hydrodynamics scheme that utilizes adaptive mesh
refinement to attain high spatial resolution.  We are interested in the interaction of
realistic galaxy clusters with their environment and therefore simulate large volumes of space and
must impose a cosmological model on our simulations.  We have chosen a flat $\Lambda$CDM cosmology
with the following parameters: $\Omega_{m} = 0.3$, $\Omega_{b} = 0.026$, $H_{0} = h \; 100 \; km \;
s^{-1} \; Mpc^{-1} = 70$ and $\sigma_{8} = 0.928$.

The simulations presented here derive from a sample of galaxy clusters from a computational volume
256 Mpc on a side.  From a coarse resolution run of the full volume we identified regions where clusters
form and rerun the simulation with the adaptive mesh framework deployed about each region of interest
in turn.  In Figure \ref{Burns:ref_vol} we show the region of space around a cluster of interest that is 
evolved with dynamic
spatial refinement.  We use from 7 to 11 levels of refinement yielding a peak spatial resolution from 16 to 1 kpc.
The dark matter particles have a mass resolution of $10^{9} h^{-1} M_{\odot}$.

The Simulated Cluster Archive, an online archive of simulated clusters of galaxies evolved in both the adiabatic 
limit and with radiative cooling, is available at http://sca.ncsa.uiuc.edu.  This archive provides interactive
tools to examine, visualize  and extract data from over 50 clusters.

\section{Radiative Cooling Simulations}
\label{Burns:rcsim}

A majority of galaxy clusters have cooling times in their central, or core, regions
that are short compared to the expected age of the cluster.   To adequately simulate
clusters,  we must account for the loss of energy from the fluid
to X-Ray radiation as it is precisely the core region that dominates the X-Ray
properties of clusters.  It is interesting, from a theoretical point of view,
to compare a sample of numerical clusters evolved with radiative cooling
only with the properties of observed clusters.
Using a large sample helps
to alleviate concerns about statistical noise in the simulated cluster properties.
We would not expect these clusters to ``look'' like clusters observed with
\textit{Chandra} in detail .   In particular there is no mechanism in the input model that
could account for the lack of gas cooler than about 1-2 keV in cluster cores as
recently reported by various authors (e.g. Sakelliou et~al. 2002).  
Nevertheless, it is useful to examine
in detail what the effect of radiative cooling is and how, specifically, the model
fails in comparison to observed clusters.  We may then take an incremental approach
among the various plausible physical mechanisms at play in cluster cores, introducing
each into the simulations in turn based on our \textit{a priori} expectations for how the
mechanism will correct the deficiencies of the previous iteration.  For example, star
formation, which will be considered later in this paper, acts as a highly
selective sink of gas \citep{motl03b}.  It removes only gas that is collapsing and rapidly cooling,
the remainder of the fluid being largely unaffected.  This property makes star
formation an interesting process to consider in removing the gas cooler than
1 keV in cluster cores.

%\begin{widetext}
   \begin{figure}
      \plotone{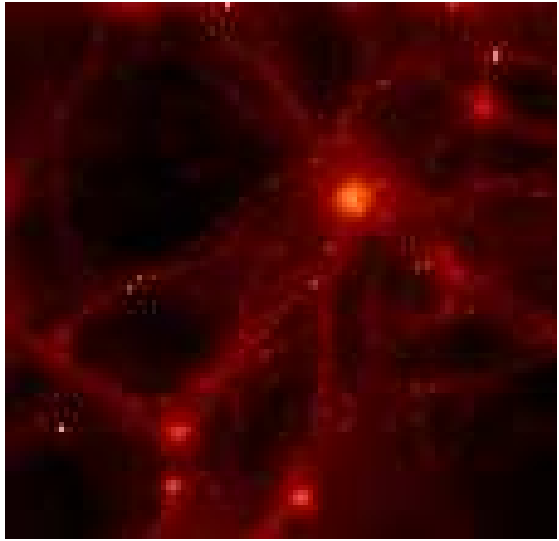}
      \figcaption{The region eligible for dynamic refinement for a typical cluster simulation.
      The region shown in the image is 36 Mpc on a side (out of the total computational volume
      with box length of 256 Mpc) while the inset highlighted region is 5 Mpc on a side.
      \label{Burns:ref_vol}}
   \end{figure}
%\end{widetext}

The loss of energy to radiation
is calculated from a tabulated cooling curve derived from a Raymond-Smith
plasma emission model assuming a constant metallicity of 0.3 relative to
solar.  The cooling curve is truncated below a temperature of $10^{4} K$.
 Every timestep, we calculate the energy radiated from each cell and
remove that amount of energy from the fluid \citep{motl03a}.    %While we do not evolve the
%cooling curve with time to account for the growing abundance of metals but
%other researchers have employed such a technique and found that it makes only
%a relatively small difference in the simulations results (reference).

   \begin{figure}
      \begin{center}
      \begin{tabular}{cc}
         \textbf{X-Ray} & \textbf{Temperature} \\
         \multicolumn{2}{c}{$\mathbf{z = 0.43}$} \\
         \begin{picture}(102,102)(0,0)
            \put(-5,51){\vector(0,-1){51}}
            \put(-15,61){\rotatebox{90}{\makebox(0,0)[tr]{5 Mpc}}}
            \put(-5,51){\vector(0,1){51}}
            \includegraphics[scale=1.0]{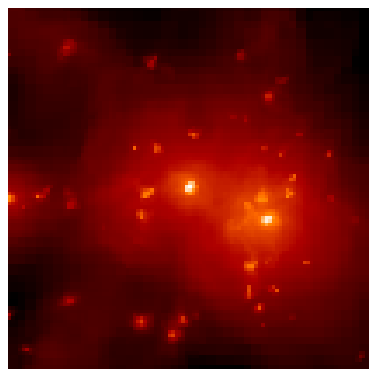}
         \end{picture} &
         \includegraphics[scale=1.0]{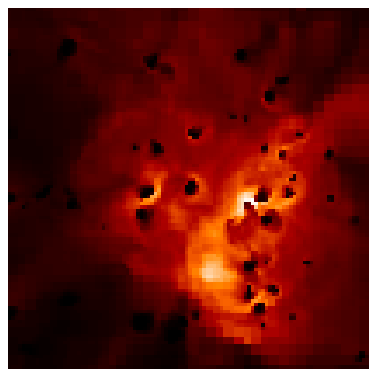} \\
         \multicolumn{2}{c}{$\mathbf{z = 0.37}$} \\
         \includegraphics[scale=1.0]{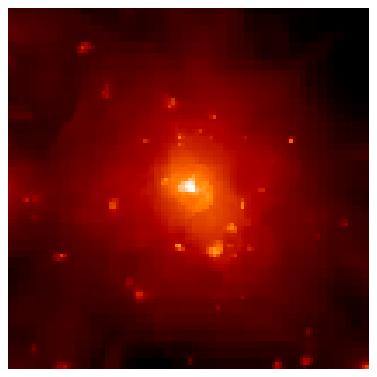} &
         \includegraphics[scale=1.0]{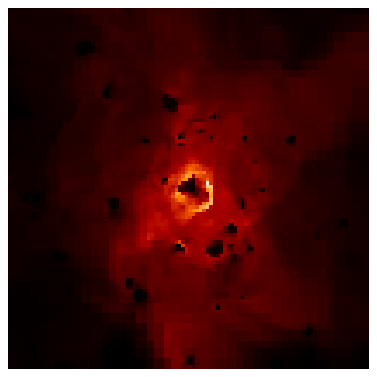} \\
         \multicolumn{2}{c}{$\mathbf{z = 0.31}$} \\
         \includegraphics[scale=1.0]{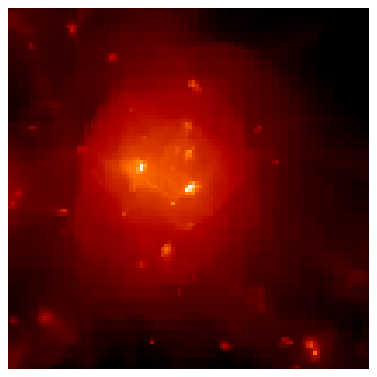} &
         \includegraphics[scale=1.0]{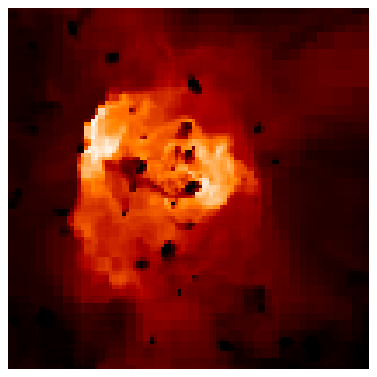} \\
         \multicolumn{2}{c}{$\mathbf{z = 0.04}$} \\
         \includegraphics[scale=1.0]{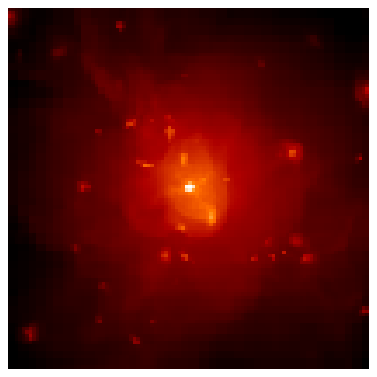} &
         \includegraphics[scale=1.0]{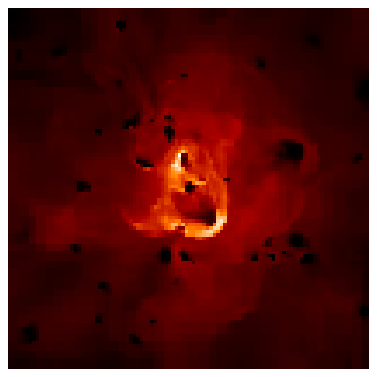} \\
         \includegraphics[scale=0.37]{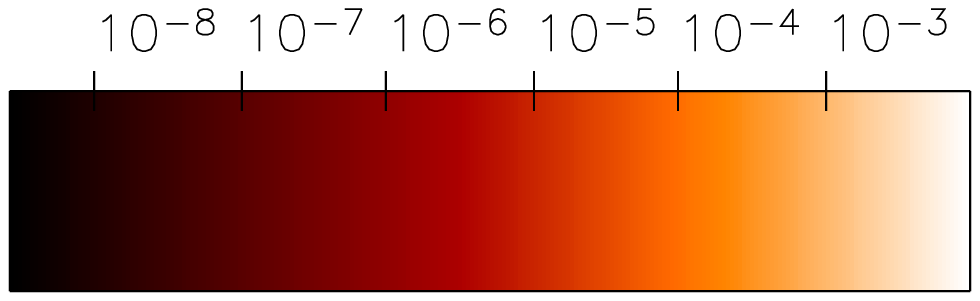} &
         \includegraphics[scale=0.37]{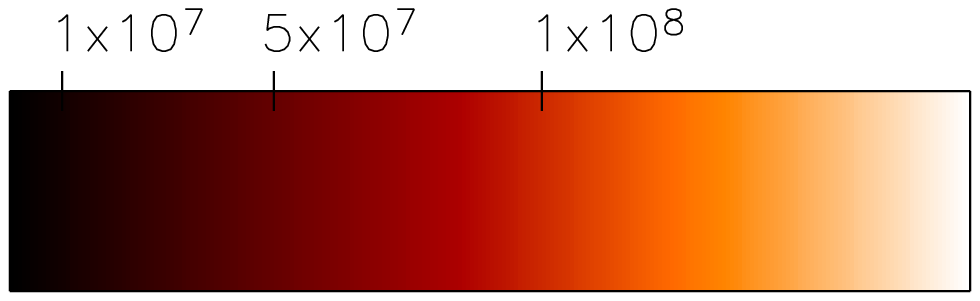} \\
      \end{tabular}
      \end{center}
      \figcaption{Three timeslices,  depicting a major merger between two approximately equal
      mass clusters, are shown in the first three rows.   The left column shows the predicted
      X-Ray surface brightness (normalized to its peak value)
      while the right column contains maps of the projected,
      emission-weighted temperature.  The timeslices at redshifts of $z = 0.43, 0.37, 0.31$
      respectively correspond to the start of the merger, closest approach of the two
      cores and finally the cores reaching their turning point.  A more typical interaction
      is shown in the bottom row, where a small subcluster (with mass of $\approx$ 10\%
      the mass of the main cluster) falls through the cluster and its core  is disrupted
      into an extended region of cooler gas behind the Mach cone of a shock front.
      \label{Burns:clrc01_col}}
   \end{figure}

\begin{center}
   \begin{figure}
      \plotone{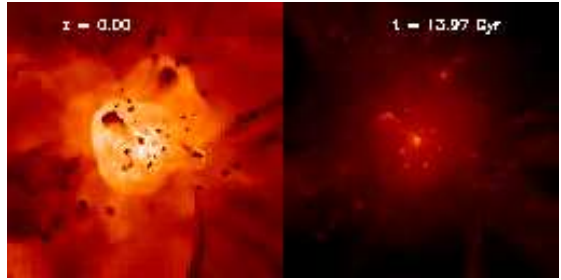}
      \figcaption{ Still frames from the end of an animation showing the evolution of a
      massive cluster ($2 \times 10^{15} \; M_{\odot}$) 
      from redshifts of 4 to the present.  The left panel shows the
      X-Ray surface brightness while the right panel shows the projected, emission-weighted
      temperature map for a region 8 Mpc on a side.  The animation
      is available at http://casa.colorado.edu/$\sim$motl/research along with other 
      movies from our simulations.  \label{Burns:clrc_mov}}
   \end{figure}
\end{center}

We have constructed a sample evolved with radiative cooling,
comprising about 75 clusters in the mass range from $4 \times 10^{14} \; M_{\odot}$
to $2 \times 10^{15} \; M_{\odot}$.  To summarize this sample, we first examine
the general appearance and typical histories of these simulated clusters.
Recall from Figure \ref{Burns:ref_vol} that we evolve the clusters along with
their surrounding environment, including the network of filaments that intersect
at nodes where clusters form.
In Figure \ref{Burns:clrc01_col} we show a sequence of images demonstrating the
interactions that are typically found in our sample clusters.  This specific cluster
is the  most massive in our sample ($M_{vir} = 2 \times 10^{15} \; 
M_{\odot}$ for an overdensity of $\delta \rho / \rho = 200$).  The cluster undergoes
a major collision between approximately equal mass components that lasts for
about $2 \; Gyr$ beginning at a redshift of $z = 0.43$ with the initial collision of
the cluster halos.  The two cores make their closest approach at a redshift of
0.37 and begin to fall back toward one another at $z = 0.31$.  Note in particular
the strong shock fronts that expand out through the clusters and the tight cores
of cooled gas behind the shocks.  The cores survive this extreme collision intact
and eventually merge into a single core at a redshift of approximately 0.25.  The 
full evolutionary sequence of this particular cluster is shown in an animation
(Figure \ref{Burns:clrc_mov}) that is available online at http://casa.colorado.edu/$\sim$motl/research.

   \begin{figure}
      \begin{center} 
      \begin{tabular}{cc}
          \textbf{Radiative Cooling} & \textbf{Adiabatic} \\
         \includegraphics[scale=1.0]{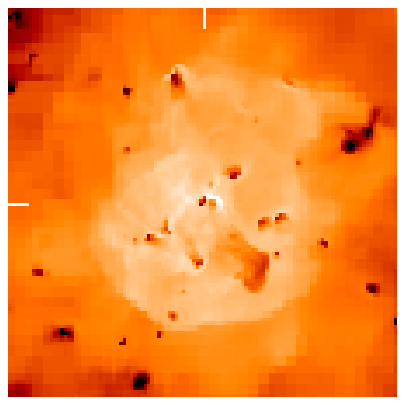} &
         \includegraphics[scale=1.0]{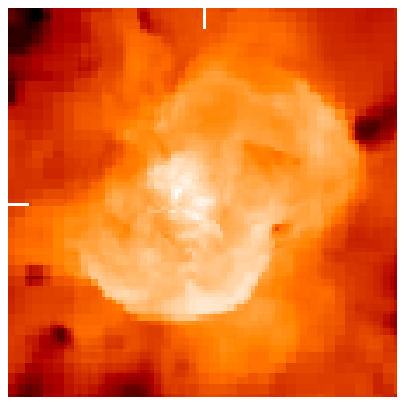} \\
         \includegraphics[scale=0.4]{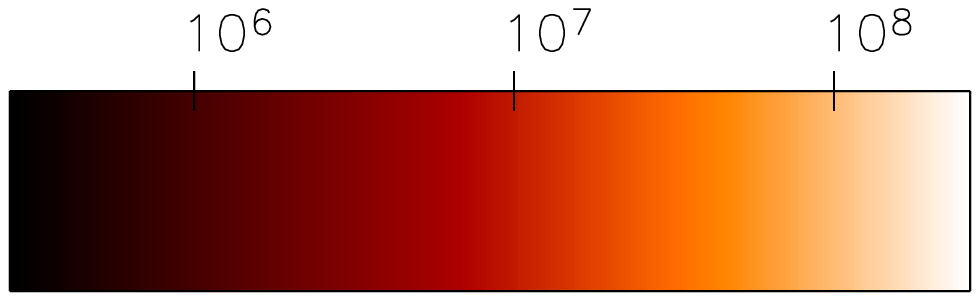} &
         \includegraphics[scale=0.4]{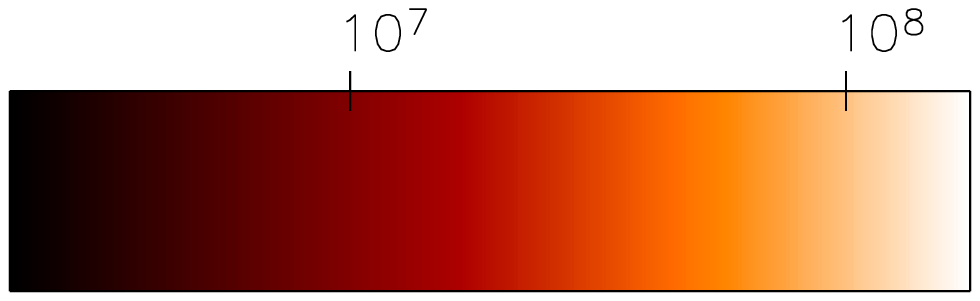} \\
         \includegraphics[scale=1.0]{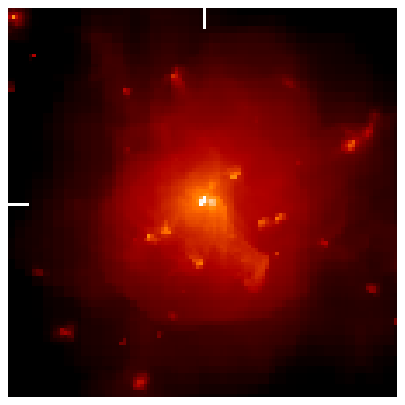} &
         \includegraphics[scale=1.0]{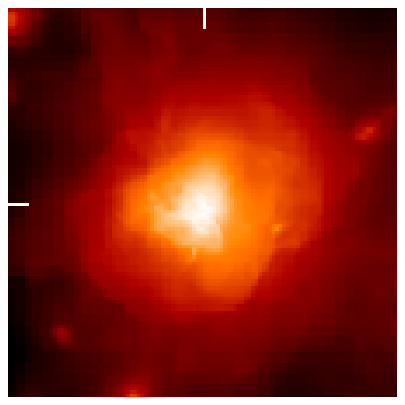} \\
         \includegraphics[scale=0.4]{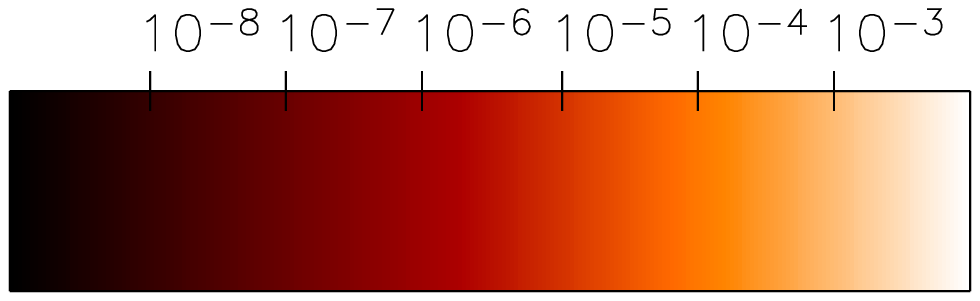} &
         \includegraphics[scale=0.4]{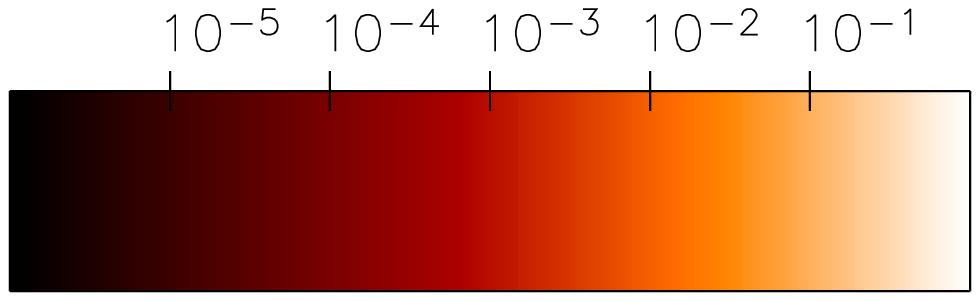} \\
      \end{tabular}
      \end{center}
      \figcaption{Comparison of radiative cooling (left column) and 
      adiabatic (right column) realizations of the same
      cluster at the present epoch. The projected, emission-weighted temperature maps 
      are shown in the top row while the bottom row contains synthetic images of the 
      X-Ray surface brightness. \label{Burns:cl01_comp}}
   \end{figure}

In addition to major collisions, most clusters in our sample experience numerous
interactions with subclusters that are less massive by a factor of 10 or more.
An example of the impact of these more frequent interactions on the appearance
of our simulated clusters is shown in the bottom panel of Figure \ref{Burns:clrc01_col}.
A small subcluster has fallen through the cluster from the upper left to bottom
right in this projection.  A strong shock front can be seen leading the subcluster's
core and this core material is being stripped to ultimately yield an extended patch
of gas that is significantly cooler than the cluster average temperature.  The 
subcluster decelerates so that the irregular patch is moving slowly compared to the 
cluster itself and the subcluster material quickly reaches  pressure equilibrium 
with the cluster potential yielding a long lived, stable structure in the cluster's 
temperature map.

   \begin{figure}
      \begin{center}
      \begin{tabular}{ccc}
         \begin{picture}(70,70)(0,0)
            \put(-5,35){\vector(0,-1){35}}
            \put(-15,47){\rotatebox{90}{\makebox(0,0)[tr]{5 Mpc}}}
            \put(-5,35){\vector(0,1){35}}
            \includegraphics[scale=1.0]{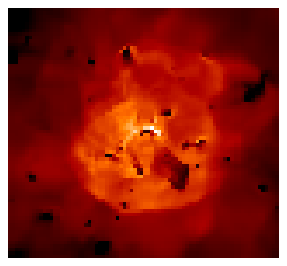}
         \end{picture} &
         \includegraphics[scale=1.0]{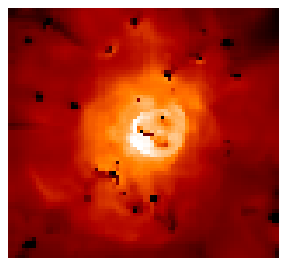} &
         \includegraphics[scale=1.0]{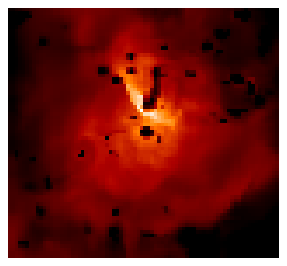} \\
         \includegraphics[scale=1.0]{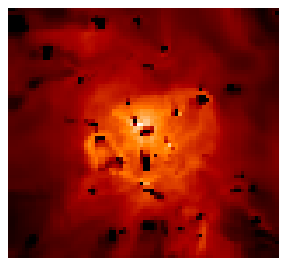} &
         \includegraphics[scale=1.0]{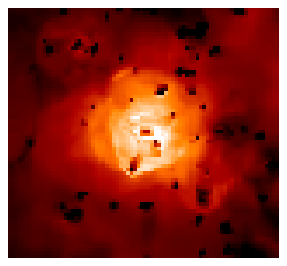} &
         \includegraphics[scale=1.0]{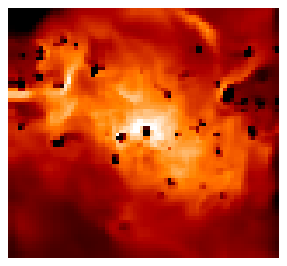} \\
         \includegraphics[scale=1.0]{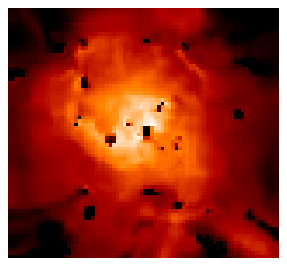} &
         \includegraphics[scale=1.0]{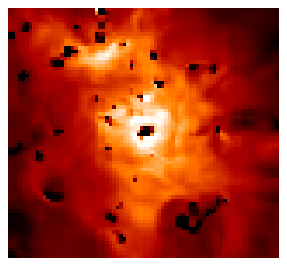} &
         \includegraphics[scale=1.0]{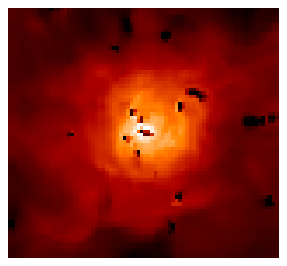} \\
         \includegraphics[scale=1.0]{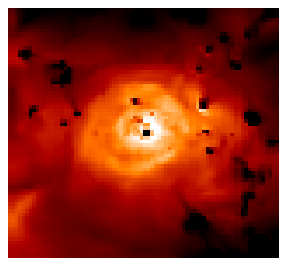} &
         \includegraphics[scale=1.0]{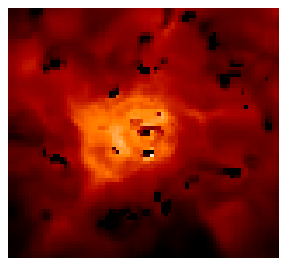} &
         \includegraphics[scale=1.0]{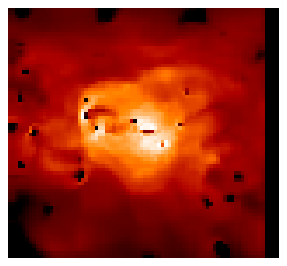} \\
         \includegraphics[scale=1.0]{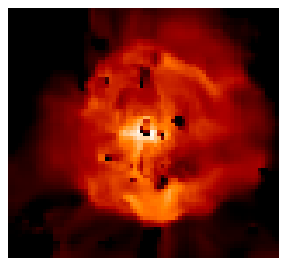} &
         \includegraphics[scale=1.0]{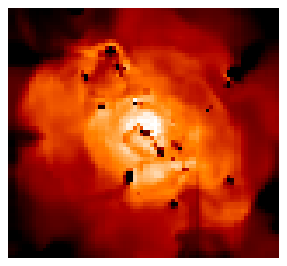} &
         \includegraphics[scale=1.0]{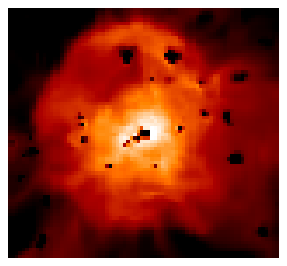} \\
         \includegraphics[scale=1.0]{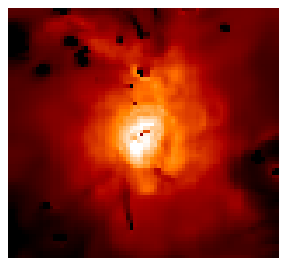} &
         \includegraphics[scale=1.0]{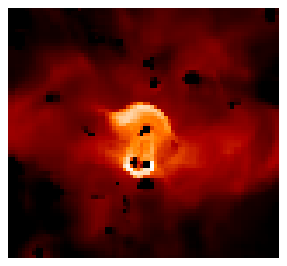} &
         \includegraphics[scale=1.0]{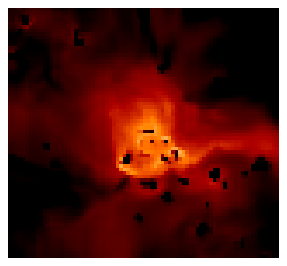} \\
         \includegraphics[scale=1.0]{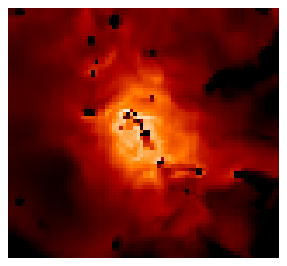} &
         \includegraphics[scale=1.0]{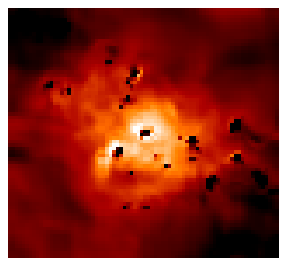} &
         \includegraphics[scale=1.0]{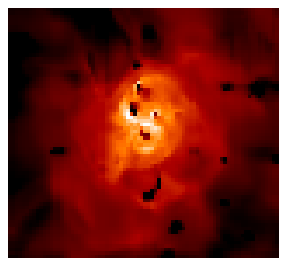} \\
         \includegraphics[scale=1.0]{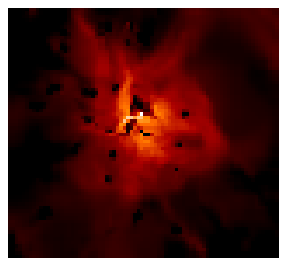} &
         \includegraphics[scale=1.0]{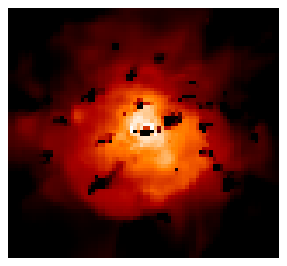} &
         \includegraphics[scale=1.0]{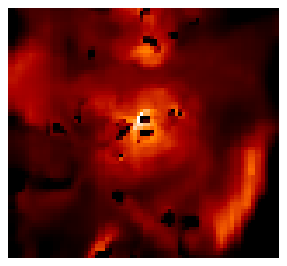} \\
      \end{tabular}
      \end{center}
      \figcaption{The projected, emission-weighted temperature maps for 24
      clusters from the sample of radiative cooling clusters.  Each image is
      5 Mpc on a side and the clusters are arranged in order of decreasing
      mass through the table.  The clusters show a wide variety of appearances
      ranging from strongly interacted to largely unperturbed and symmetric.
      \label{Burns:clrc_samp}}
   \end{figure}
%\end{widetext}

In Figure \ref{Burns:cl01_comp} we show X-Ray surface brightness and temperature
images from the same cluster evolved with radiative cooling and in the adiabatic
limit to highlight the impact of radiative cooling.  Both simulations begin from
identical initial conditions which means that the merger history is nearly identical
for both clusters.  The most dramatic difference between the two realizations is 
the presence of numerous cool cores when radiative cooling is introduced.  These cores
are dense (bright in the surface brightness map) and cool (dark in the temperature
map) and significantly alter the cluster appearance on scales comparable to a
typical core radius.  In the adiabatic
limit, these substructures quickly mix with the main cluster during interactions.
With the addition of radiative cooling, the cool core in each substructure becomes
a long lived entity that can survive a passage through the cluster.
On larger scales, the  temperature and density are in fact similar
between the two simulations, for example, examine the curved shock front below
the cluster center in both realizations.  To lowest order, radiative cooling has
an impact only in dense regions where the cooling time is short.  To the next order,
radiative cooling changes the interaction of subclusters within the cluster which results
in strong, localized shocks and trails of stripped, cooler gas from the in-falling subcluster
cores.

The range of structures in the radiative cooling clusters are shown in Figure
\ref{Burns:clrc_samp} where we plot the projected, emission-weighted temperature
maps at the present epoch from a subset of 24 clusters.  The clusters are arranged
in order of decreasing mass through the table with the most massive cluster
($2 \times 10^{15} \; M_{\odot}$) in the upper left corner.  Each image is centered on
the cluster center of mass.  The central cool cores vary widely in their strength (size)
and shape (ranging from nearly spherical to bar-like to irregular filaments that connect
multiple cores).  Many of the clusters contain significant amounts of shocked
gas from ongoing interactions and, in general,  a large number of cool cores from
previous interactions.

A typical cluster evolution from a redshift of 4 to the present day has been prepared
as an animation.  The still frames from the end of the animation are shown in Figure
\ref{Burns:clrc_mov} where the projected, emission-weighted temperature and X-Ray
surface brightness are shown for a region $8 \; Mpc$ on a side.  The animation from
this particular cluster evolution and other simulations are available online from
http://casa.colorado.edu/$\sim$motl/research.  This  cluster gains most of its
mass from one major collision (see also Figure \ref{Burns:clrc01_col} which shows
additional images from this particular cluster) and undergoes numerous other interactions.

\subsection{What Cooling Does}
\label{Burns:cool}

With radiative cooling, we have seen that subcluster cores often merge with the
existing core.  This observation suggests a \textbf{new, hierarchical formation
scenario for cool cores} where cores are built along with the cluster itself
through the hierarchical structure formation process of the standard, cold dark matter
model.

This hypothesis suggests that rich clusters with strong cool cores (those clusters
with a large central surface brightness excess and strongly declining temperature
profiles in their cores) should be found in dense supercluster environments as
these are the regions with the most cool fuel for the cluster core.  Such a
relationship has been found previously by \citet{lok99}.

\begin{figure}
      \begin{center}
      \begin{tabular}{cc}
         \textbf{X-Ray} & \textbf{Temperature} \\
         \\ 
         \begin{picture}(114,114)(0,0)
            \put(-5,57){\vector(0,-1){57.0}}
            \put(-15,70){\rotatebox{90}{\makebox(0,0)[tr]{5 Mpc}}}
            \put(-5,57){\vector(0,1){57.0}}
            \includegraphics[scale=1.0]{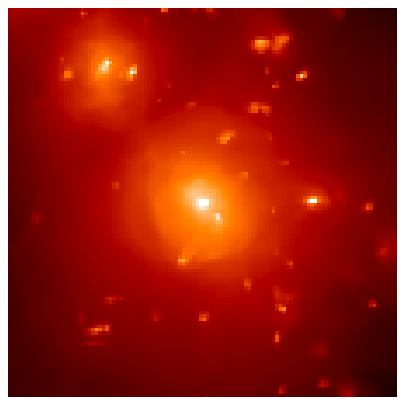}
         \end{picture} &
         \begin{picture}(114,114)(0,0)
            \includegraphics[scale=1.0]{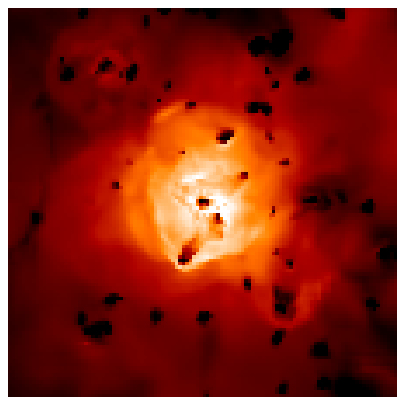}
         \end{picture} \\
         \includegraphics[scale=0.4]{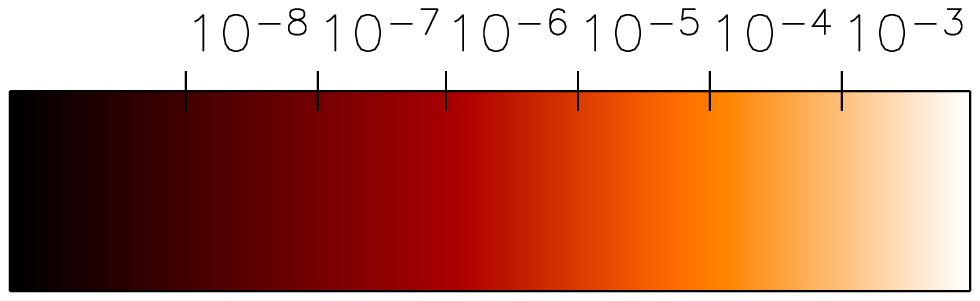} &
         \includegraphics[scale=0.4]{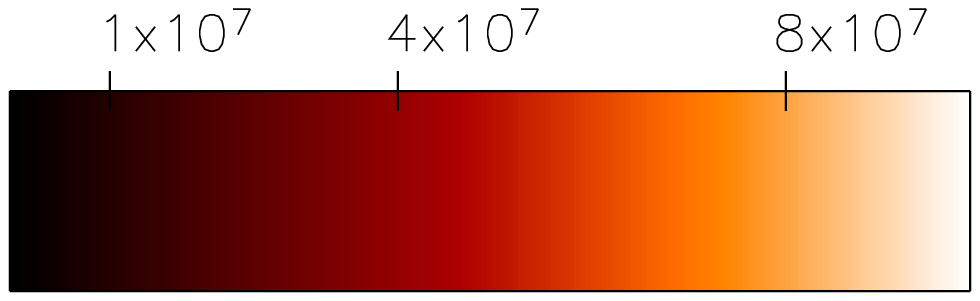} \\
      \end{tabular}
      \end{center}
      \figcaption{An example of the rich array of substructure present in the radiative
      cooling cluster sample at the present epoch.  The X-Ray surface brightness and projected,
      emission-weighted temperature maps show numerous cool cores and the interaction of
      an in-falling subcluster with the main cluster.  \label{Burns:clrc_subst}}
   \end{figure}

Significantly, the hierarchical formation scenario for cool cores naturally accounts for the rich
substructure seen with high-resolution X-Ray instruments such as \textit{Chandra}
and \textit{XMM}.  An example case of this rich substructure is shown in Figure
\ref{Burns:clrc_subst}.  Predicted X-Ray surface brightness and projected,
emission-weighted temperature maps at the present epoch are shown for this
relatively high mass cluster ($M_{vir} \approx 1 \times 10^{15} \; M_{\odot}$).
The surface brightness is normalized to its peak value and please note the
extreme contrast in the X-Ray image.  If cast in terms of a typical observation,
the rich substructure would largely be absent in the X-Ray.  It is important to
note that the complex and irregular temperature structure would be relatively
easy to detect provided enough counts could be detected to extract and fit
spectra for different spatial regions.

\begin{figure}
      \begin{center}
      \begin{tabular}{cc}
         \textbf{X-Ray} & \textbf{Temperature} \\
         \\
         \begin{picture}(114,114)(0,0)
            \put(-5,57){\vector(0,-1){57}}
            \put(-15,70){\rotatebox{90}{\makebox(0,0)[tr]{5 Mpc}}}
            \put(-5,57){\vector(0,1){57}}
            \includegraphics[scale=1.0]{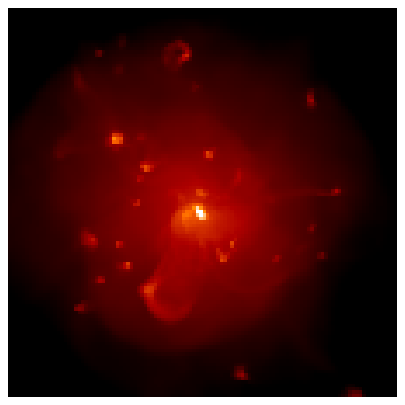}
         \end{picture} &
         \begin{picture}(114,114)(0,0)
            \includegraphics[scale=1.0]{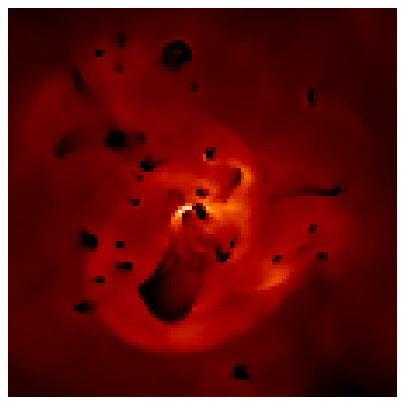}
         \end{picture} \\
         \includegraphics[scale=0.4]{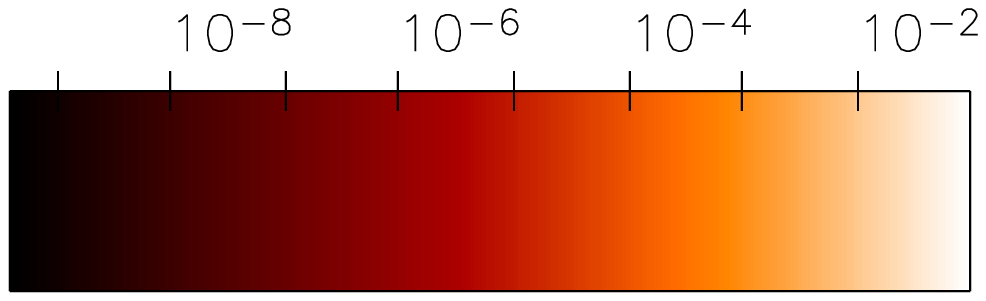} &
         \includegraphics[scale=0.4]{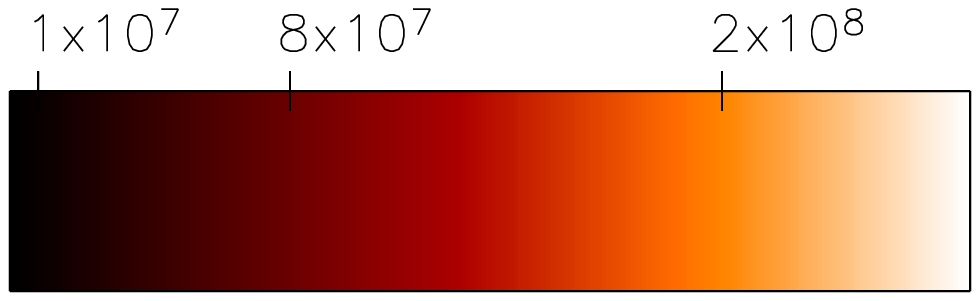} \\
      \end{tabular}
      \end{center}
      \figcaption{Demonstration of the formation of cold fronts in clusters of galaxies.
      A subcluster has fallen through the main cluster from the upper right and its gas
      has been stripped resulting in the sharp discontinuity separating cool and hotter
      gas to the lower left of the cluster center.  The subcluster has decelerated and
      the leading shock front is no longer visible.  Instead, there is now a relatively
      long-lived patch of cooler gas.  \label{Burns:clrc_cf}}
   \end{figure}

Many clusters have been found to possess extremely complex temperature
distributions, even in cases when the X-Ray emission appears to arise
from a symmetric, relaxed cluster.  In some instances, the combination
of the gas density and temperature indicates the presence of contact
discontinuities in the central region of the cluster.  Termed ``Cold
Fronts'', these features arise naturally from cluster mergers.  An example
from our simulations with radiative cooling only is shown in Figure
\ref{Burns:clrc_cf}.  A small subcluster with mass $\approx 0.1$ times
the cluster mass has fallen through the cluster and the leading shock
front has dissipated.  Tidal forces and ram pressure stripping both
act to disrupt the subcluster core and reshape the material into
an extended patch of cooler gas.  Since the virial temperature of
subclusters will be significantly lower than that of the cluster, 
a similar process will act even in the adiabatic limit \citep{evr02}
as the subcluster material is stripped from its dark matter halo and
expands to reach pressure equilibrium with the cluster gas.

\begin{figure}
      \begin{center}
      \begin{tabular}{cc}
         \multicolumn{2}{c}{\textbf{Filaments - Abell 1795}} \\
         \begin{picture}(128,128)(0,0)
            \includegraphics[scale=1.0]{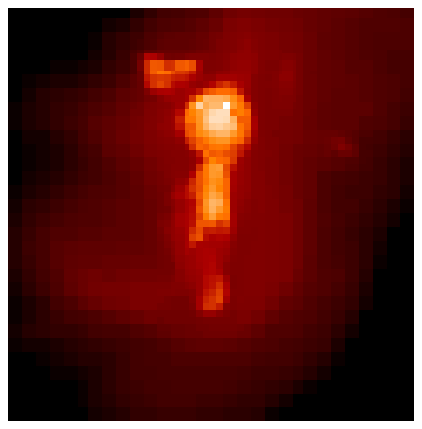}
         \end{picture} &
         \begin{picture}(128,128)(0,0)
            \includegraphics[scale=1.0]{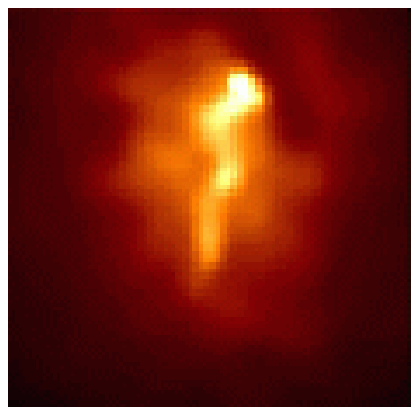}
         \end{picture} \\
         \multicolumn{2}{c}{\textbf{Cold Fronts - Abell 2256}} \\
         \begin{picture}(128,128)(0,0)
            \includegraphics[scale=1.0]{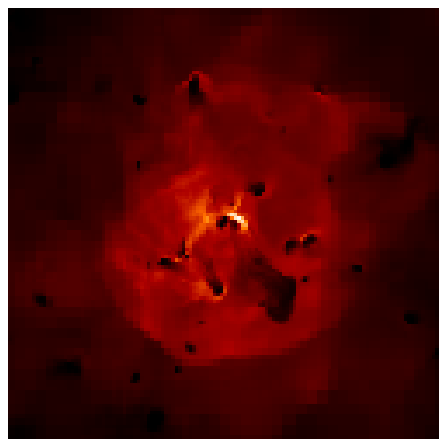}
         \end{picture} &
         \begin{picture}(128,128)(0,0)
            \includegraphics[scale=1.0]{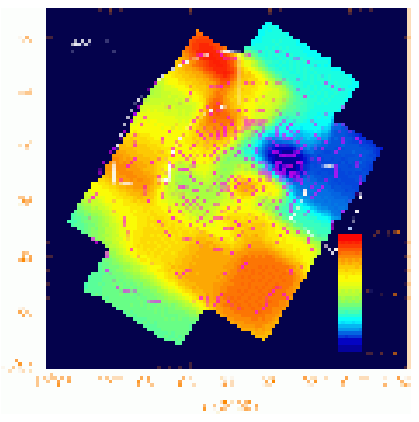}
         \end{picture} \\
         \multicolumn{2}{c}{\textbf{Colliding Bullets - 1E0657-56}} \\
         \begin{picture}(128,128)(0,0)
            \includegraphics[scale=1.0]{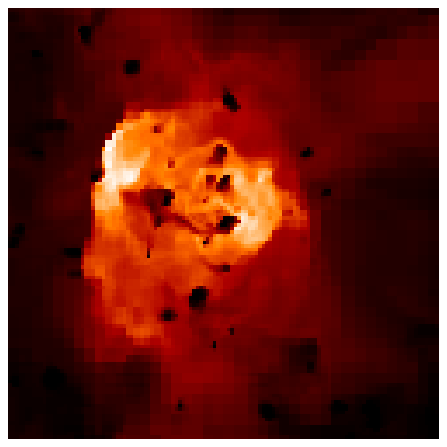}
         \end{picture} &
         \begin{picture}(128,128)(0,0)
            \includegraphics[scale=1.0]{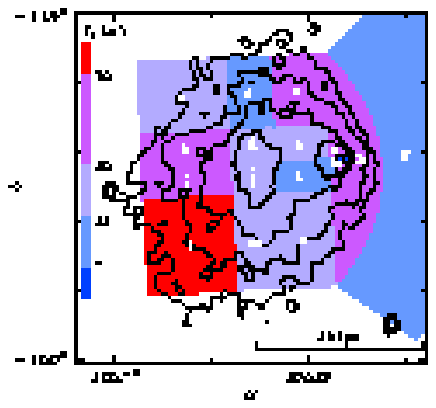}
         \end{picture} \\
      \end{tabular}
      \end{center}
      \figcaption{Comparison of features from cluster simulations (left) with recent observations from
      \textit{Chandra} (right).  In the right top row, we show the filament of cool gas trailing behind the
       core in Abell 1795 \citep{fab01}.  Complex, filamentary structures like this are common
       in our simulations and indicate the relative flow of material past the core.  Many clusters,
       such as Abell 2256 \citep{sun02}, show evidence for ``Cold Fronts'' when examined with
       high-resolution instruments such as \textit{Chandra}.   Similar, non-uniform temperature
       distributions are common features in our simulations.  Finally, in the bottom row we show
       the high redshift cluster 1E0657-56 which, as reported by \citet{mark02}, shows the collision
       between two clusters and a shock front centered on the small cool core that corresponds to
       a Mach number of approximately 2.  For comparison, we show a major collision with similar
       Mach number between two approximately equal mass subclusters. \label{Burns:obs_ex}}
   \end{figure}

A table comparing snapshots from our simulations and recent X-Ray
observations with \textit{Chandra} is shown in Figure \ref{Burns:obs_ex}.
There is a strong commonality between these observed features, including
filaments, cold fronts and strong shocks from cluster-cluster collisions, 
and the structures seen in our simulations.  As clusters form through
hierarchical mergers, a process that continues to this day, the relatively 
X-Ray bright tracer material from cool cores is subjected to a complex flow through the
cluster which results in a rich array of morphological features.

\begin{center}
   \begin{figure}
      \plotone{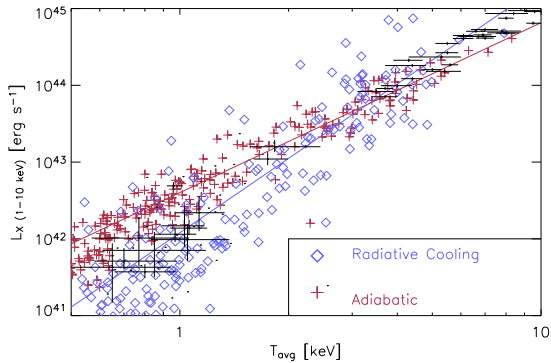}
      \figcaption{The relation between X-Ray luminosity and temperature for clusters from
      both the adiabatic and radiative cooling samples.  The data points with error
      bars hotter than 2 keV are from \citet{mark98} while the cooler cluster and groups
      are a random subset from the recent atlas of groups from \citet{mul03}.  For the simulated
      clusters, the X-Ray luminosity is measured in the 1-10 keV energy band from all material
      within the virial radius of the cluster (taken to be the radius for an overdensity
      $\delta \rho / \rho$ of 200) and the average cluster temperature is measured from the
      emission-weighted temperature within one half of a virial radius.  The solid, colored
      lines show the least-squares fit relation for the 100 most massive objects in each sample.
      The adiabatic sample has a best fit slope of $\approx 2$ while the radiative cooling sample
      has a steeper relation with a slope of $\approx 3$. \label{Burns:lx_t}}
   \end{figure}
\end{center}

Radiative cooling enforces a scale length on halos, the relatively high
density regions of cluster cores have short cooling times, and this
breaks the self-similar scaling of cluster properties that would otherwise
be present.  In Figure \ref{Burns:lx_t} we show the $L_{X}$ \textit{vs}
$T$ relation for clusters from both our adiabatic and radiative cooling
samples.  The adiabatic clusters do indeed exhibit an $L_{X} - T$ relation
consistent with the self-similar scaling law.  The radiative cooling
clusters obey a steeper relation, closer to what has been observed, 
with the hottest and most massive clusters being over-luminous compared
to the self-similar result and the cooler clusters and groups being
under-luminous.

\begin{center}
   \begin{figure}
      \plotone{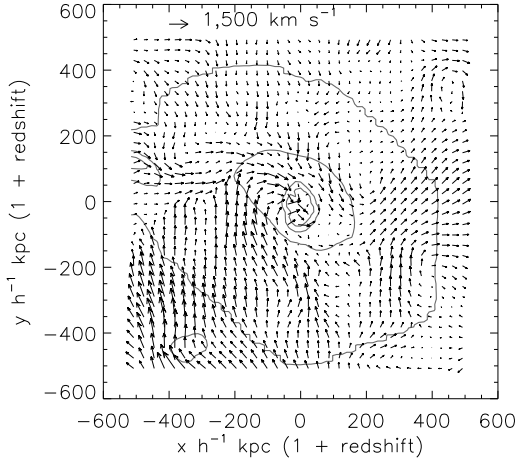} 
      \figcaption{A two-dimensional slice of the velocity field from one
       particular cluster during a period free from major interactions.
       The contours are of the projected  X-Ray emissivity.  Note the
       large scale bulk flows through the cluster and the rotation of fluid
       in the core region.  \label{Burns:vel}}
   \end{figure}
\end{center}

An important question about our simulations is, ``What is the fate of the
cooling gas?''.  If an idealized cluster were left in isolation and its
atmosphere were allowed to cool it would inevitably form a ``cooling
flow'' in its center.  As the gas cools it loses pressure 
support and must flow inward in an attempt to regain hydrostatic
equilibrium with the dark matter potential well.  The inflow pushes
the gas to a higher density which only shortens the cooling time
further.  The solution is thus forced to be an inflow with more and more
gas collapsing into the center of the cluster.

This scenario applies only to a cluster in isolation; recall from Figure
\ref{Burns:ref_vol} that we are considering clusters as they form in the
larger cosmological environment.  The filaments channel material into the
cluster and impose complex flow boundary conditions that episodically bring
discrete subclusters into the  cluster in addition to the continuous inflow
of warm, low density gas.  The resulting flow field within clusters is
very complex and the typical speeds are far in excess of the flow speeds
expected to arise from the onset of a simple ``cooling flow'' model.  An
example velocity field from one of our clusters is shown in Figure
\ref{Burns:vel} which depicts a slice through the velocity field in the
central Mpc of the cluster.  The last major collision for this cluster
was approximately 2 Gyr in the past and from the contours in the X-Ray
emission we see that cluster appears fairly relaxed.  The isophotes are elliptical
and there is twisting of the isophotes as one moves into the core region
but there is no strong evidence for recent activity from the X-Ray
data alone.  When one examines the flow field, however, one sees that
there are significant inflows from the lower and upper left corners
of the frame, an outflow to the right and rotation as the in-falling
material flows through the core.  This flow corresponds to speeds often
in excess of $1,000 \; km \; s^{-1}$.  Similar patterns are present in
this particular cluster at other epochs and for other slices and for the
remaining clusters in our sample as well.

\subsection{Problems and What Radiative Cooling Alone Can Not Explain}
\label{Burns:cool_prob}

We have seen that our simulations with radiative cooling have led to the
hypothesis that cool cores in clusters of galaxies are formed through
hierarchical mergers along with the cluster itself.  As a consequence,
we expect a variety of morphological features indicative of mergers to
be present even in clusters with significant cool cores.  These cool
cores do not correspond to the picture of a cooling flow model, simply
because of the cosmological infall of material into the cluster which
enforces complex boundary conditions on the flow field.  To summarize
briefly, the cores contain cool gas but this gas is not at rest and
is certainly not isolated from its environment.

The radiative cooling model can not be the last word in simulations
of clusters, however.  The model suffers from some serious deficiencies
that must be addressed with more physically complete simulations.  First,
the radiative cooling only model over-produces cool cores and these cores
are too robust and resilient during collisions and mergers.  Nearly every
dark matter potential well in our simulations is occupied by a halo of 
gas that has cooled significantly.  As these halos merge and form clusters
at the present epoch, the individual cores often survive and maintain an
easily discernible identity independent of the cluster.  While this is
necessary to explain the irregular arrangement of gas in different temperature
phases in clusters, the radiative cooling only model is too efficient
in this respect as compared to observations.

   \begin{figure}
      \begin{center}
      \begin{tabular}{c}
         \textbf{Temperature} \\
         \begin{picture}(233,96)(0,0)
            \put(-5,48){\vector(0,-1){48}}
            \put(-15,60){\rotatebox{90}{\makebox(0,0)[tr]{9 Mpc}}}
            \put(-5,48){\vector(0,1){48}}
            \includegraphics[scale=1.0]{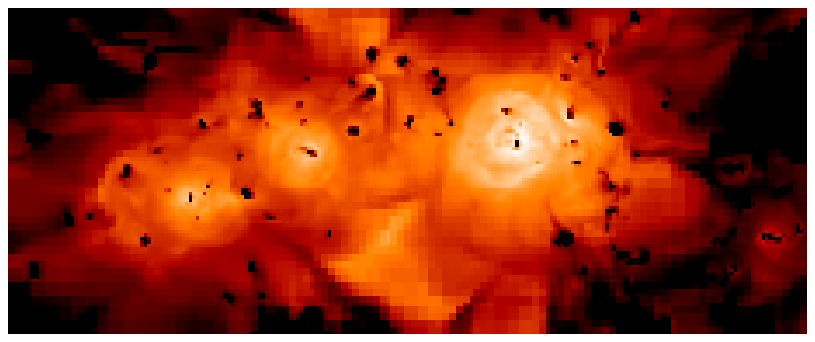}
         \end{picture} \\
         \textbf{Dark Matter Density} \\
         \begin{picture}(233,96)(0,0)
            \includegraphics[scale=1.0]{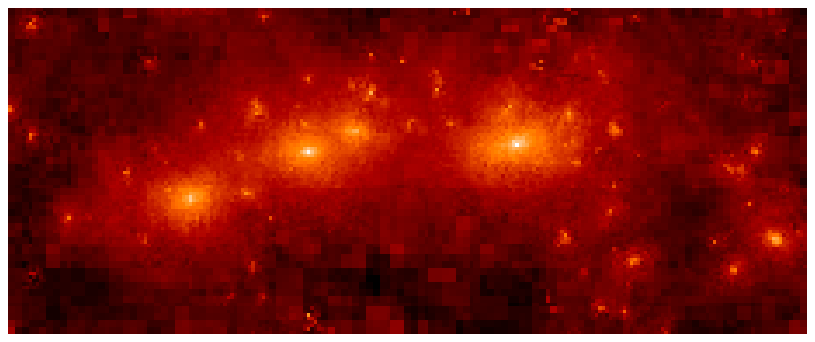}
         \end{picture} \\
      \end{tabular}
      \end{center}
      \figcaption{Projections of the emission-weighted temperature and dark matter
      particles from a rich supercluster region in our simulation volume.  Essentially
      all potential wells formed by the dark matter particles are occupied by cores
      of cool gas in the radiative cooling simulations. \label{Burns:clrc09}}
   \end{figure}

   \begin{figure}
      \begin{center}
      \plotone{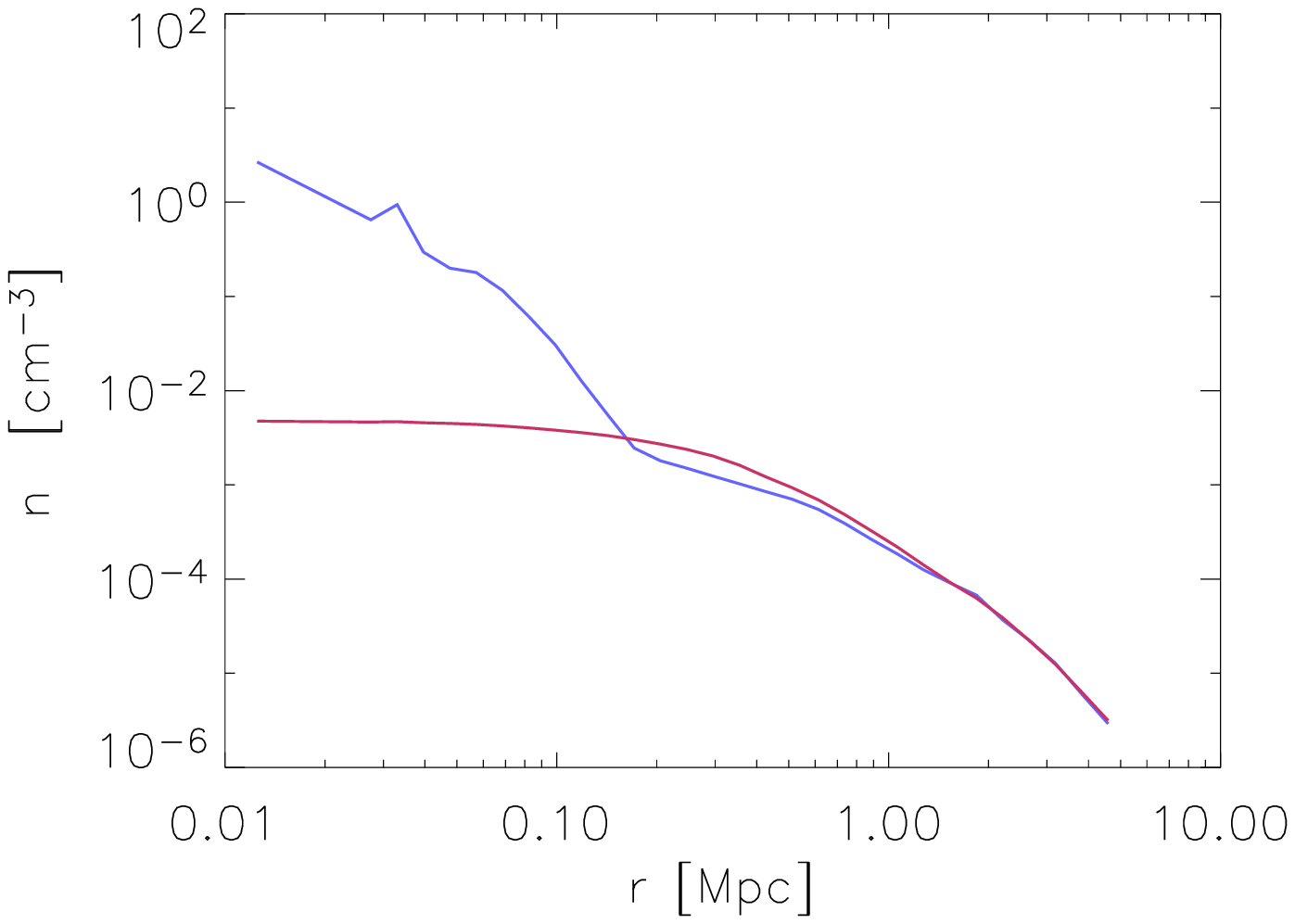}
      \plotone{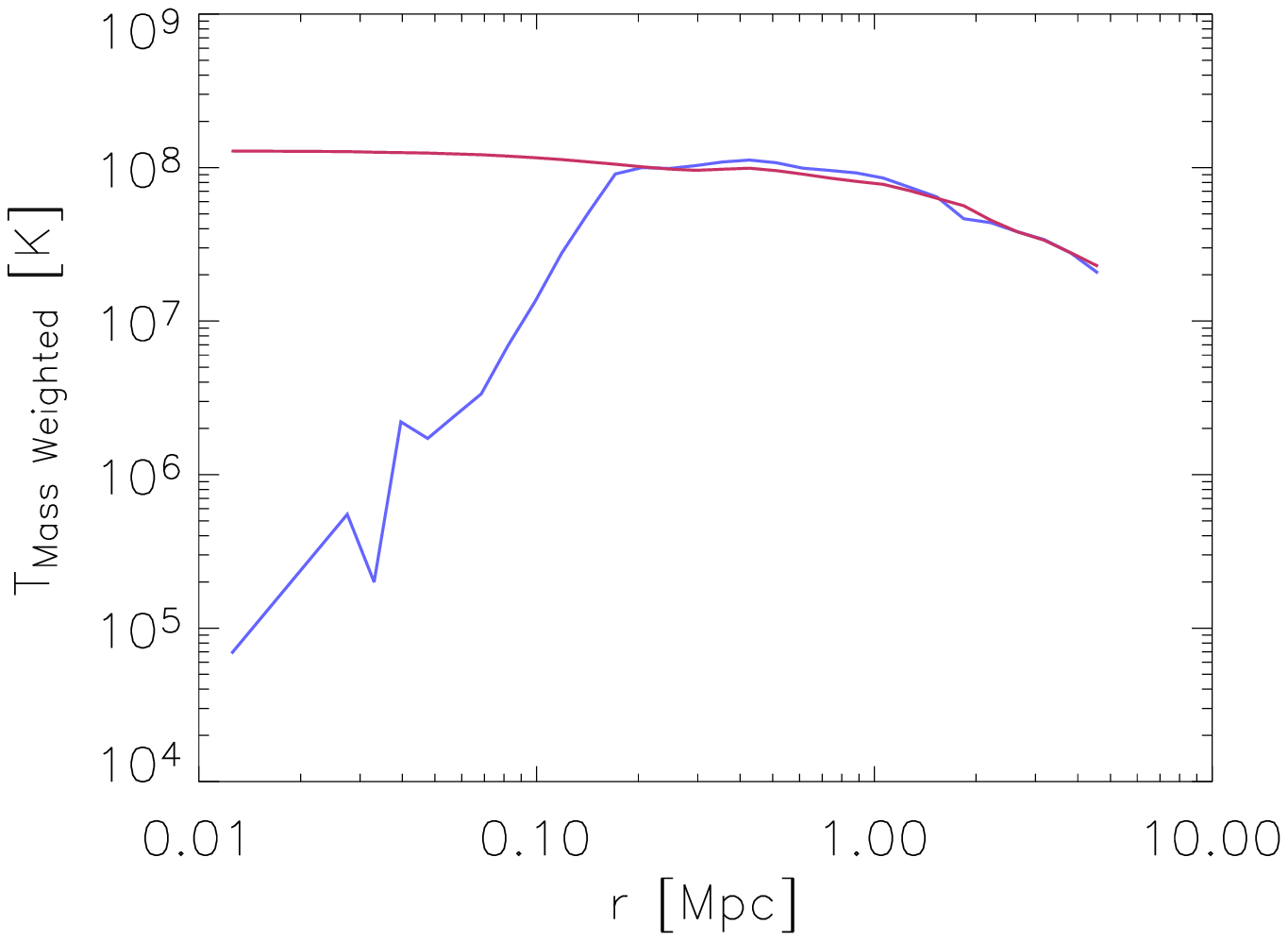}
      \end{center}
      \figcaption{Spherically averaged profiles of the gas density and mass-weighted
      temperature from the same cluster evolved in the adiabatic limit (red curves)
      and with radiative cooling (blue curves).  The cool core is evidenced as the
      cool, dense gas within the central 100 kpc. \label{Burns:sph_prof}}
   \end{figure}

For example, in Figure \ref{Burns:clrc09} we show the projected, emission-weighted
temperature map and the corresponding projection of the dark matter particles
for a large area containing a rich supercluster.   There is an obvious correspondence
between each dark matter halo and a steep decline in the gas temperature.  The
images also have many examples of interactions, none of which destroy the cool cores.

Why are the cool cores so resilient?  With radiative cooling, the central density of
the gas has been increased significantly as depicted in Figure \ref{Burns:sph_prof}.
The cores are therefore akin to rigid billiard balls sloshing through the cluster
medium.  The steep density contrast shelters the cores from ram pressure stripping
as subclusters rain in.  As can also be seen in Figure \ref{Burns:sph_prof}, the
temperature structure of the core gas contains material that is significantly cooler
than $1 \; keV$.  This core material, if present in real clusters, would create spectral
features in the UV and X-Ray that are not observed in any cluster.  

A more complete computational model for galaxy clusters must address the following
points.  The cores must be softened in that they must be made less dense and therefore
more susceptible to disruption during interactions.  While a majority of clusters
harbor cool cores there are definite examples of clusters that do not.  With the
correct input physics, a computational model for clusters will reproduce the observed
abundance of cool cores at any epoch.  A more daunting challenge is to account for the
lack of gas at intermediate temperatures ( $1 \; keV < T < 0.1 \; keV$ ) in cluster
cores.

\section{Star Formation}
\label{Burns:sf}

Star formation should provide a natural mechanism to soften cool 
cores by both transforming dense, rapidly cooling gas into stars 
and also by heating the surrounding material with the energy injected 
from supernova explosions. 
We have incorporated the
star formation prescription from Cen and Ostriker (1992) to introduce some of
the effects of star formation in our simulations.  The code examines all grid 
cells at the finest refinement level above a specified overdensity threshold.  According 
to this star formation recipe, fluid is converted to collisionless ``star'' particles 
if the following conditions are met:
\begin{itemize}
   \item The fluid is undergoing compression ( $\mathbf{\nabla} \cdot 
         \mathbf{v} < 0$ )
   \item Rapid cooling (local $t_{cool} < t_{dyn}$ )
   \item The mass of fluid in the cell ( $m_{b}$ ) exceeds the Jean's mass
\end{itemize}
When all conditions are met, a new star particle is created with a mass
$m_{b} \; \eta \; \Delta t / t_{dyn}$ and this amount of mass is removed from the
fluid.  The star formation rate is thus coupled to the local dynamical
time, $t_{dyn}$, while $\eta$ parameterizes the efficiency of star formation
and $\Delta t$ is the simulation timestep increment.

Once formed, the new star particle begins to deposit energy in the fluid
to simulate the explosion of prompt,  type II supernovae.  The strength of supernova
feedback is controlled by an efficiency, $\epsilon_{feedback}$, which gives the
thermal energy injected by the star particle in terms of the particle's
rest mass energy.  The star particle also pollutes the fluid with a 
passive metallicity tracer field. 

   \begin{figure}
      \begin{center}
      \begin{tabular}{cc}
         \textbf{X-Ray} & \textbf{Temperature} \\
         \multicolumn{2}{c}{$\mathbf{ \epsilon_{feedback} = 10^{-4}}$} \\
         \begin{picture}(110,110)(0,0)
            \put(-5,55){\vector(0,-1){55}}
            \put(-15,65){\rotatebox{90}{\makebox(0,0)[tr]{5 Mpc}}}
            \put(-5,55){\vector(0,1){55}}
            \includegraphics[scale=1.0]{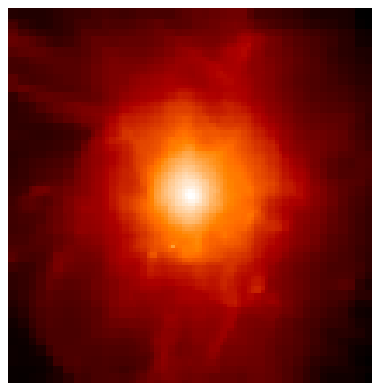}
         \end{picture} &
         \begin{picture}(110,110)(0,0)
            \includegraphics[scale=1.0]{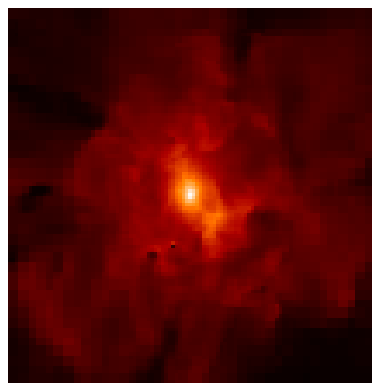}
         \end{picture} \\
         \multicolumn{2}{c}{$\mathbf{ \epsilon_{feedback} = 10^{-6}}$} \\
         \begin{picture}(110,110)(0,0)
            \includegraphics[scale=1.0]{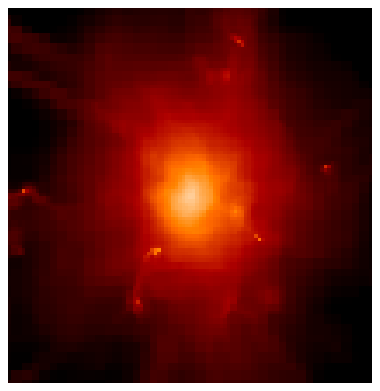}
         \end{picture} &
         \begin{picture}(110,110)(0,0)
            \includegraphics[scale=1.0]{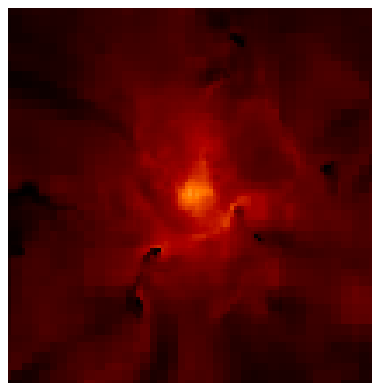}
         \end{picture} \\
         \multicolumn{2}{c}{$\mathbf{ \epsilon_{feedback} = 0}$} \\
         \begin{picture}(110,110)(0,0)
            \includegraphics[scale=1.0]{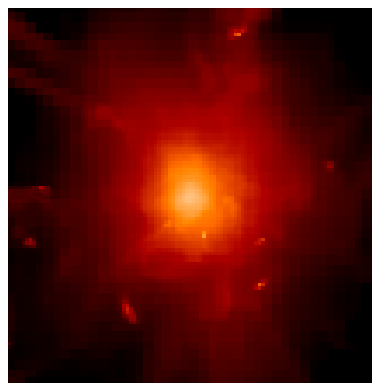}
         \end{picture} &
         \begin{picture}(110,110)(0,0)
            \includegraphics[scale=1.0]{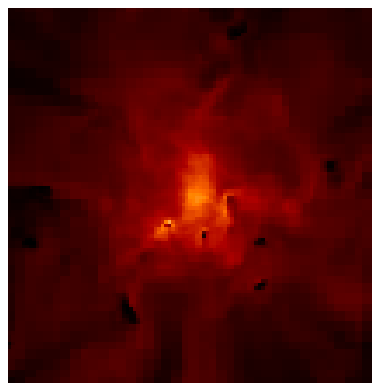}
         \end{picture} \\
         \includegraphics[scale=0.4]{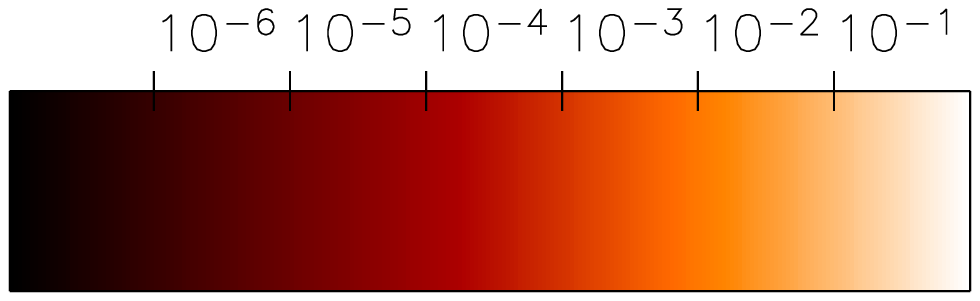} &
         \includegraphics[scale=0.4]{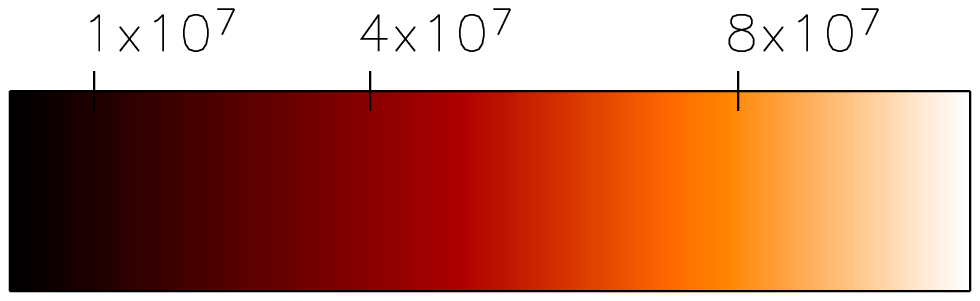} \\
      \end{tabular}
      \figcaption{Comparison of a moderate mass cluster evolved with star formation and
      varying strength of thermal feedback, $\epsilon_{feedback}$, from an unrealistically
      large value in the top row, to a moderate value in the middle row, to the limiting
      case of no feedback
      in the bottom row.  As the feedback strength is lowered, subclusters have stronger
      cool cores.  The amount of gas converted to stars increases as the feedback strength
      is lowered, ranging from $~ 2\%$ to $20\%$ to $33\%$ for the case of zero feedback.
      The strength of feedback
      serves as a throttle on star formation and one generation of stars can disrupt (heat)
      gas in their local environment, preventing the formation of future generations.
      \label{Burns:clsf_feedback}}
      \end{center}
   \end{figure}

In our initial work with star formation, we explore the parameter space of
the star formation model and assume that star formation continues to the
present epoch.

The dominant parameter in the star formation model is the strength of
supernova feedback.  This parameter controls the condition of the
fluid in star forming regions and largely dictates the amount of baryons
converted into stars.  In Figure \ref{Burns:clsf_feedback} we show the
end states for a test cluster evolved with three  different values for
the strength of supernova feedback (from top to bottom in Figure \ref{Burns:clsf_feedback}
the strength varies from $\epsilon_{feedback} = 10^{-4}, 4 \times 10^{-6}, 0$).

The impact of the remaining model parameters is less important.  The overdensity
threshold where star formation will be considered can impact when star formation
begins in the simulation and the efficiency of star formation, $\eta$, will 
impact the amount of mass converted into stars, but to a lesser extent than the
feedback strength.  

It should be noted that none of the parameters significantly impact the clusters 
that ultimately
form from the star-forming halos.  The resulting clusters
tend to be similar to their adiabatic realizations, there are no cool cores in any
cluster, and the clusters obey self-similar scaling relations.  To explain the
absence of cool cores in simulations with continuous star formation, consider the
limiting case where the strength of feedback has been set to zero.  This is the
case where all cooling baryons form stars and any material that would result in
a cool core is simply removed from the gas phase.  As the strength of feedback is
increased, fewer and fewer baryons are transformed into stars. This is
because gas is again removed from the cool phase but this time by heating from supernova
in a previous generation of stars.  The net effect is that, when star formation operates
continuously throughout the simulation, any fluid element that could become a cool core
is quickly removed from the fluid.

\section{Truncated Star Formation}
\label{Burns:tsf}

The star formation model that we employ is simplistic in many respects, not the
least of which is the fact that we must approximate a complex physical process 
operating on mass scales around a solar mass with a collective process
operating on scales up to a billion times larger.  The star formation prescription
is also free-form in that you can not \textit{a priori} impose the amount
of fluid converted to stars or the magnitude of the star formation rate with
time in the simulation.

\begin{figure}
      \begin{center}
      \plotone{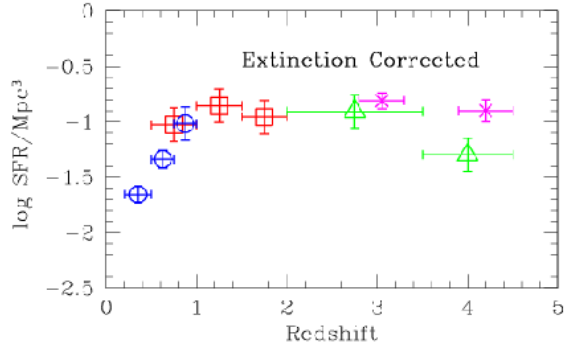}
      \figcaption{The inferred star formation rate from \citet{ste99} after correction for extinction
       by dust.  The star formation rate is consistent with a constant rate up to redshifts of
       approximately 1 and a subsequent decline in more recent time.\label{Burns:steidel}}
       \end{center}
   \end{figure}

However, it is well known that the star formation rate has declined in the interval from
a redshift of one to the present and was likely approximately constant before
that time (see for instance, Figure \ref{Burns:steidel}, where we reproduce the
cosmological history of the star formation rate compiled by \citet{ste99}).
As a constraint on the simulations with star formation we may simply halt
star formation at a given epoch, $z_{truncation}$ and complete the simulation
with radiative cooling only.  This approach essentially approximates the curve
in Figure \ref{Burns:steidel} with a step function.  

While this approach is rather crude, there clearly must be a value for 
$z_{truncation} = z_{crit}$ where a given cluster has just enough cool gas to form a cool
core at the present epoch.  If star formation were allowed to continue on
beyond $z_{crit}$, proto-clusters would be depleted of their cool gas and
the resulting cluster would essentially be adiabatic.  If star formation
is halted before $z_{crit}$, cool cores would begin to be overabundant just
as they are in the radiative cooling only simulations.

\begin{figure}
      \begin{center}
      \begin{tabular}{cc}
         \textbf{X-Ray} & \textbf{Temperature} \\
         \multicolumn{2}{c}{\textbf{Adiabatic}} \\
         \begin{picture}(90,90)(0,0)
            \put(-5,44){\vector(0,-1){44}}
            \put(-15,55){\rotatebox{90}{\makebox(0,0)[tr]{5 Mpc}}}
            \put(-5,44){\vector(0,1){44}}
            \includegraphics[scale=1.0]{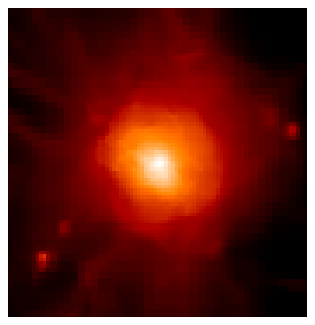}
         \end{picture} &
         \begin{picture}(90,90)(0,0)
            \includegraphics[scale=1.0]{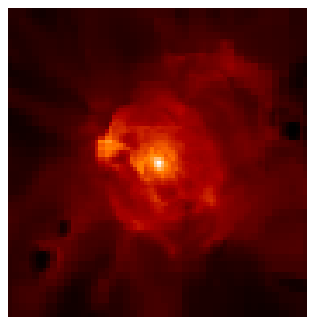}
         \end{picture} \\
         \multicolumn{2}{c}{$\mathbf{z_{truncation} = 0}$} \\
         \begin{picture}(90,90)(0,0)
            \includegraphics[scale=1.0]{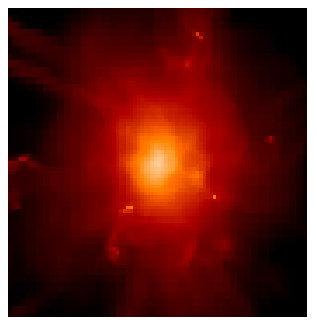}
         \end{picture} &
         \begin{picture}(90,90)(0,0)
            \includegraphics[scale=1.0]{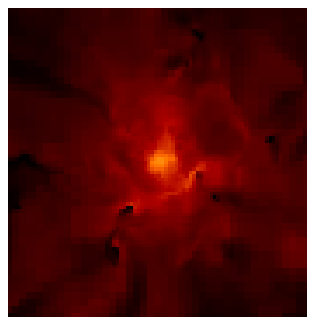}
         \end{picture} \\
         \multicolumn{2}{c}{$\mathbf{z_{truncation} = 1}$} \\
         \begin{picture}(90,90)(0,0)
            \includegraphics[scale=1.0]{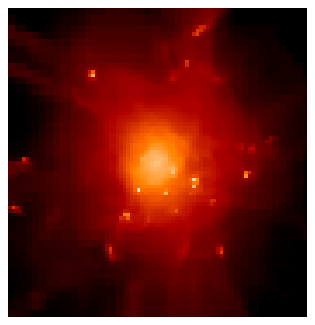}
         \end{picture} &
         \begin{picture}(90,90)(0,0)
            \includegraphics[scale=1.0]{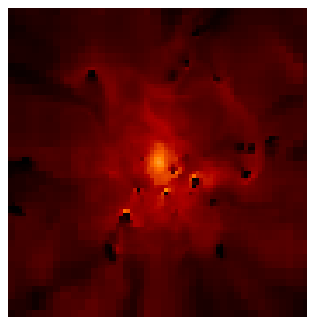}
         \end{picture} \\
         \multicolumn{2}{c}{$\mathbf{z_{truncation} = 1.5}$} \\
         \begin{picture}(90,90)(0,0)
            \includegraphics[scale=1.0]{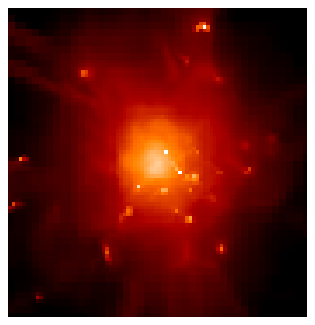}
         \end{picture} &
         \begin{picture}(90,90)(0,0)
            \includegraphics[scale=1.0]{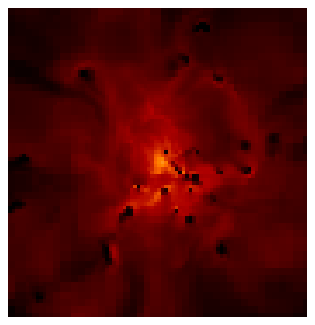}
         \end{picture} \\
         \multicolumn{2}{c}{$\mathbf{z_{truncation} = 2}$} \\
         \begin{picture}(90,90)(0,0)
            \includegraphics[scale=1.0]{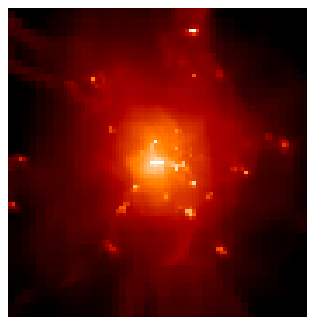}
         \end{picture} &
         \begin{picture}(90,90)(0,0)
            \includegraphics[scale=1.0]{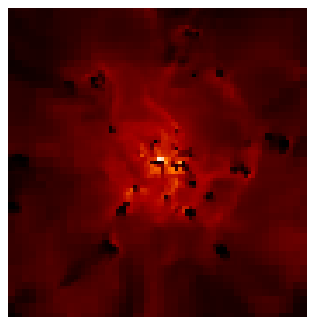}
         \end{picture} \\
         \multicolumn{2}{c}{\textbf{Cooling}} \\
         \begin{picture}(90,90)(0,0)
            \includegraphics[scale=1.0]{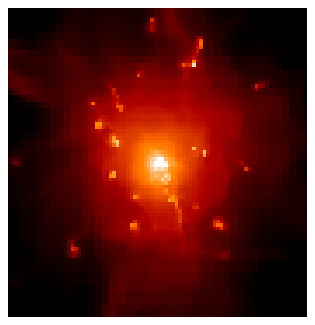}
         \end{picture} &
         \begin{picture}(90,90)(0,0)
            \includegraphics[scale=1.0]{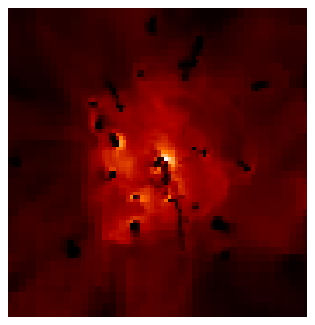}
         \end{picture} \\
         \includegraphics[scale=0.32]{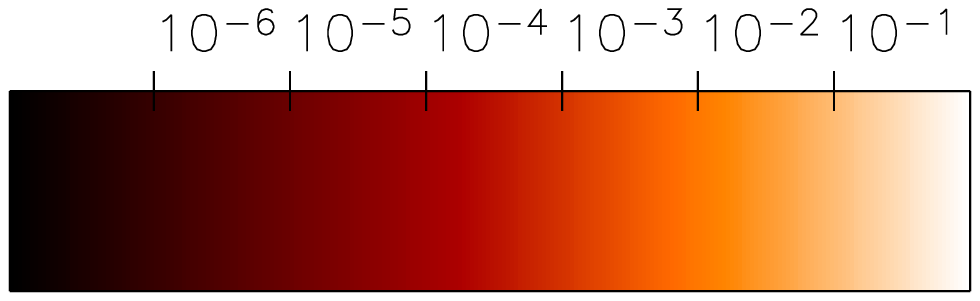} &
         \includegraphics[scale=0.32]{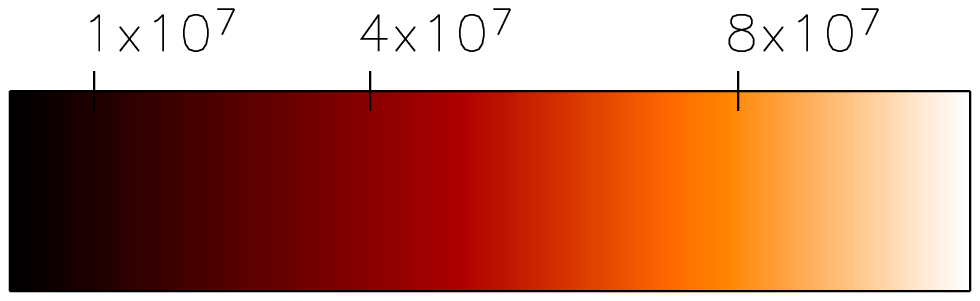} \\
      \end{tabular}
      \end{center}
      \figcaption{The X-Ray surface brightness and projected, emission-weighted
      temperature maps from the same cluster as in Figure \ref{Burns:clsf_feedback}
      showing the Adiabatic, Radiative Cooling realizations as well as for several
      values of $z_{truncation}$.  As star formation is halted at progressively
      higher redshifts, subclusters have stronger cool cores up to the Radiative
      Cooling simulation where no gas is lost to star formation.  \label{Burns:clsf_trunc}}
   \end{figure}

\begin{figure}
      \begin{center}
      \plotone{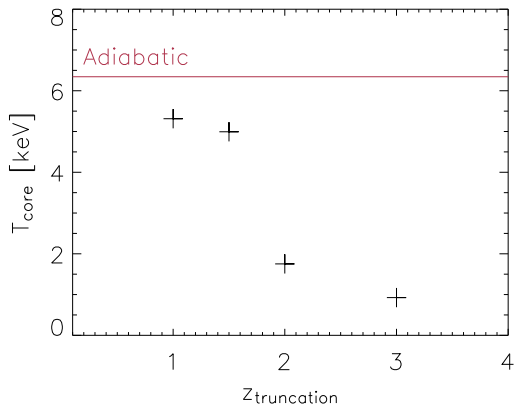}
      \figcaption{The core temperature from the same cluster evolved with star
      formation truncated at the indicated redshift.  The core temperature is
      measured from the projected, emission-weighted temperature within a 50
      kpc region centered on the system center of mass. For reference, the core
      temperature from the adiabatic simulation is also shown.  This particular
      cluster undergoes an abrupt transition from having a cool core for values
      of $z_{truncation} \ge 2$ and having no cool core for $z_{truncation} \le
      1.5$ (see also Figure \ref{Burns:clsf_trunc} for images of the clusters)
      \label{Burns:trunc_temp}}
      \end{center}
   \end{figure}

In Figures \ref{Burns:clsf_trunc} and \ref{Burns:trunc_temp} we show results from
a numerical experiment where we simulated an otherwise identical cluster with
star formation truncated at a set of values for   $z_{truncation}$.  In Figure
\ref{Burns:clsf_trunc} we can see the progression of the cluster's
appearance from the smooth, adiabatic realization to the highly structured cooling
only realization as star formation is halted at higher and higher redshifts.
From this experiment, this cluster has $1.5 < z_{crit} < 2$.

This result is emphasized in Figure \ref{Burns:trunc_temp} where the core temperature
of the cluster is plotted against the value of $z_{truncation}$.  The core temperature
jumps abruptly from about 2 to 5 keV as the cool core disappears in this particular
cluster.

It is encouraging that, for this limited experiment, the value of $z_{crit}$ is reasonable
in light of the results in Figure \ref{Burns:steidel}.  First, the model is very simple
and star formation does not decay, it is abruptly halted.  Second, the results from Figure
\ref{Burns:steidel} at different epochs come from very different environments.  At high
redshifts, the star formation rate is indeed measured from the regions of high
overdensity that likely will become clusters at the present day.  For the lower redshift
results, where the star formation rate is falling, these are measured in the dynamically
younger field environment, not in clusters.  It is therefore reasonable for the 
value of $z_{crit}$ to exceed the value inferred for the field in the nearby universe.

\section{Conclusions}
\label{Burns:conc}

To summarize our results for simulations with radiative cooling only, we find that the
model of radiative cooling only can explain:
\begin{itemize}
   \item The wide array of structures seen in recent X-Ray observations, including cool
         cores, filaments and rich substructure
   \item Formation and evolution of cool core clusters via accretion from supercluster
         filaments
   \item Agreement with observed cluster scaling relations such as the $L_{X}$ \textit{vs}
         $T$ relationship
   \item Presence of cool cores but no inflows (no ``cooling flows'') are present.
\end{itemize}

However, the radiative cooling only model can not be complete.  Outstanding problems
with the cooling only model include:
\begin{itemize}
   \item Overabundance of cool cores (all clusters have cool cores).
   \item Core temperatures are $< 1 \; keV$.
\end{itemize}

When star formation is introduced into our simulations, cool gas is rapidly channeled into
stars leaving no gas to form the cool cores present in a majority of present day clusters.
One promising approach in our simulations is to truncate star formation at a specific
redshift as a means of approximating the observed decline in the universal star formation
rate.  We have found that star formation truncated at $z < 2$ produces:
\begin{itemize}
   \item Softer (low density) cool cores
   \item Core temperatures at observed levels ($T_{core} \approx 1-2 \; keV$).
\end{itemize}

%% Figures may be included in a single column using the "figure" environment.
%% Multi-column figures may be included using the "figure*" environment. The
%% AASTeX "plotone" and "plottwo" should be used to insert the postscript
%% files.
%\begin{figure}
%\plotone{fig1.ps}
%\figcaption{This is the caption of a figure.
%\label{Scott:myplot}}
%\end{figure}

%%%%%%%%%%%%%%%%%%%%%%%%%%%%%%%%%%%%%%%%%%%%%%%%%%%%%%%%%%%%%%%%%%%%%%%%%%%%%%%%
%%
%% This is the end of the main part of the article
%%
%%%%%%%%%%%%%%%%%%%%%%%%%%%%%%%%%%%%%%%%%%%%%%%%%%%%%%%%%%%%%%%%%%%%%%%%%%%%%%%%

\acknowledgements
%Funding sources should be acknowledged here.

This research was partially supported by grant TM3-4008A from NASA.
The simulations presented were conducted on
the Origin2000 system at the National Center for Supercomputing Applications
at the University of Illinois, Urbana-Champaign through computer allocation
grant AST010014N.

%%%%%%%%%%%%%%%%%%%%%%%%%%%%%%%%%%%%%%%%%%%%%%%%%%%%%%%%%%%%%%%%%%%%%%%%%%%%%%%%
%%
%% Your list of references starts here.  If you use BibTeX, please
%% include with your submission the .bbl file generated by LaTeX.
%% For BibTeX users, please use the apj.bst file, available on the
%% conference website. e.g.
%%
%%\bibliography{bibfile}

\begin{thebibliography}{}

\bibitem[{{Bialek} {et~al.}(2002){Bialek}, {Evrard}, \& {Mohr}}]{evr02}
{Bialek}, J.~J., {Evrard}, A.~E., \& Mohr, J.~J. 2002, \apj, 578, 9

\bibitem[{{Fabian} {et~al.}(2001){Fabian}, {Sanders}, {Ettori}, {Taylor}, {Allen}, {Crawford}, {Iwasawa}, \& {Johnstone}}]{fab01}
{Fabian}, A.~C., {Sanders}, J.~S., {Ettori}, S., {Taylor}, G.~B., {Allen}, S.~W., {Crawford}, C.~S., {Iwasawa}, K., \& {Johnstone}, R.~M. 2001, \mnras, 321, L33

\bibitem[{{Loken} {et~al.}(1999){Loken}, {Melott}, \& {Miller}}]{lok99}
{Loken}, C., {Melott}, A., \& {Miller}, C.~J. 1999, \apj, 520, 5

\bibitem[{{Markevitch} (1998){Markevitch}}]{mark98}
{Markevitch}, M. 1998, \apj, 504, 27

\bibitem[{{Markevitch} {et~al.}(2002){Markevitch}, {Gonzalez}, {David}, {Vikhlinin}, {Murray}, {Forman}, {Jones}, /& {Tucker}}]{mark02}
{Markevitch}, M., {Gonzalez}, A.~H., {David}, L., {Vikhlinin}, A., {Murray}, S., {Forman}, W., {Jones}, C., \& {Tucker}, W. 2002, \apj, 567, L27

\bibitem[{{Motl} {et~al.}(2003a){Motl}, {Burns}, {Loken}, {Bryan}, \& {Norman}}]{motl03a}
{Motl}, P.~M., {Burns}, J.~O., {Loken}, C., {Bryan}, G.~L., \& {Norman}, M.~L. 2003, \apj, \textit{in press}

\bibitem[{{Motl} {et~al.}(2003b){Motl}, {Burns}, {Bryan}, \& {Norman}}]{motl03b}
{Motl}, P.~M., {Burns}, J.~O., {Bryan}, G.~L., \& {Norman}, M.~L., 2003, Proceedings of \textit{The
Riddle of Cooling Flows in Galaxies and Clusters of Galaxies}

\bibitem[{{Mulchaey} {et~al.}(2003){Mulchaey}, {David}, {Mushotzky}, \& {Burstein}}]{mul03}
{Mulchaey}, J.~S., {David}, D.~S., {Mushotzky}, R.~F., \& {Burstein}, D. 2003, \apjs, 146, 353

\bibitem[{{Sakelliou} {et~al.}(2002){Sakelliou}, {Peterson}, {Tamura}, {Paerels}, {Kaastra}, {Belsole}, {B\"{o}hringer}, {Branduardi-Raymont}, {Ferrigno}, {den Herder}, {Kennea}, {Mushotzky}, {Vestrand}, \& {Worrall}}]{sak02}
{Sakelliou}, I., {Peterson}, J.~R., {Tamura}, T., {Paerels}, F.~B.~S., {Kaastra}, J.~S., {Belsole}, E., {B\"{o}hringer}, H., {Branduardi-Raymont}, G., {Ferrigno}, C., {den Herder}, J.~W., {Kennea}, J., {Mushotzky}, R.~F., {Vestrand}, W.~T., \& {Worrall}, D.~M. 2002, \aap, 391, 903

\bibitem[{{Steidel} {et~al.}(1999){Steidel}, {Adelberger}, {Giavalisco}, {Dickinson}, \& {Pettini}}]{ste99}
{Steidel}, C.~C., {Adelberger}, K.~L., {Giavalisco}, M., {Dickinson}, M., \& {Pettini}, M. 1999, \apj, 519, 1

\bibitem[{{Sun}, {et~al.}(2002){Sun}, {Markevitch}, \& {Vikhlinin}}]{sun02}
{Sun}, M., {Markevitch}, M., \& {Vikhlinin}, A. 2002, \apj, 565, 867

%\bibitem[{{Abell} {et~al.}(1989){Abell}, {Corwin}, \& {Olowin}}]{aco89}
%{Abell}, G.~O., {Corwin}, H.~G., \& {Olowin}, R.~P. 1989, \apjs, 70, 1

%\bibitem[{{Sarazin}(1988)}]{sar88}
%{Sarazin}, C.~L. 1988, X-ray Emissions from Clusters of Galaxies
%(Cambridge: Cambridge Univ. Press)

\end{thebibliography}
%%\bibliographystyle{apj}
%%
%% where bibfile.bib is your list of references in BibTeX format.
%% Please use the "thebibliography" evironment rather than the
%% "references" environment.
%%
%%%%%%%%%%%%%%%%%%%%%%%%%%%%%%%%%%%%%%%%%%%%%%%%%%%%%%%%%%%%%%%%%%%%%%%%%%%%%%%%

\end{document}